\documentclass[final,5p,times,twocolumn]{elsarticle}
\usepackage{lineno,hyperref,amsmath,amsthm,amssymb,amsfonts,ragged2e,color,subfig}
\journal{``High Temperature"}
\begin{document}
\begin{frontmatter}
\title{Electrostatic dust-acoustic rogue waves in an electron depleted dusty plasma}
\author{J. N. Sikta$^{*,1,2}$, N. A. Chowdhury$^{**,2,3}$, A. Mannan$^{2,4}$, S. Sultana$^{2,5}$, and A. A. Mamun$^{2}$}
\address{$^1$ Faculty of Science and Information Technology, Daffodil International University, Dhaka-1207, Bangladesh\\
$^2$ Department of Physics, Jahangirnagar University, Savar, Dhaka-1342, Bangladesh\\
$^3$ Plasma Physics Division, Atomic Energy Centre, Dhaka-1000, Bangladesh\\
$^4$ Institut f\"{u}r Mathematik, Martin Luther Universit\"{a}t Halle-Wittenberg, Halle, Germany\\
$^5$ Fakult\"{a}t f\"{u}r Physik und Astronomie, Ruhr-Universitt\"{a}t Bochum, D-44780 Bochum, Germany\\
E-mail: $^{*}$jebunphy@gmail.com, $^{**}$nurealam1743phy@gmail.com}
\begin{abstract}
The formation of the gigantic dust-acoustic rouge waves (DARWs) in an electron
depleted unmagnetized opposite polarity dusty plasma system is theoretically
observed for the first time. The nonlinear Schr\"{o}dinger equation (derived by
utilizing the reductive perturbation method) has been analytically as well as
numerically analyzed to identify the basic features (viz., height, thickness,
and modulational instability, etc.) of DARWs. The results obtained form this
investigation should be useful in understanding the basic properties of these
rouge waves which can predict to be formed in electron depleted unmagnetized opposite
polarity dusty plasma systems like mesosphere, F-rings of Saturn, and
cometary atmosphere, etc.
\end{abstract}
\end{frontmatter}
\section{Introduction}
\label{1sec:Introduction}
Opposite polarity (OP) dusty plasma (OPDP) is demonstrated as fully ionized gases consisting of massive  negatively and positively
charged dust grains as well as ions and electrons in presence
of electrostatic and gravitational force fields, and is identified in space, viz., cometary
tails \cite{Hossen2016a}, interstellar clouds \cite{Hossen2016b,Hossen2017}, Planetary rings \cite{Shukla1992},
solar system \cite{Hossen2017}, Earth polar mesosphere \cite{Hossen2016a}, magnetosphere of
Jupiter \cite{Hossen2016a}, and laboratory situations, viz., laser-matter interaction
\cite{Shahmansouri2013}. First, Rao \textit{et al.} \cite{Rao1990} traced that how the presence
of the  massive dust grains alters the picture of the dynamics of dusty plasma medium (DPM), and also theoretically
predicted new kind of low-frequency dust-acoustic (DA) waves (DAWs). The distinction of low-frequency
DAWs from ion-acoustic waves (IAWs) and experimental verification was, finally, confirmed by
Barkan \textit{et al.} \cite{Barkan1995} in DPM. The experimental identification
of the DAWs in DPM have mesmerized many plasma physicists to investigate numerous modern electrostatic eigen modes,
viz., dust-drift waves \cite{Shukla1991}, DA solitary waves (DA-SWs) \cite{Shahmansouri2013},
DAWs \cite{Hossen2016a,Hossen2016b,Hossen2017}, dust lattice waves \cite{Melandso1996},
DA shock waves (DA-SHWs) \cite{Ferdousi2015}, and dust-ion-acoustic waves (DIAWs) \cite{Shukla1992}
to understand the basic features of the various nonlinear electrostatic structures associated with
the propagation of low frequency electrostatic disturbance.

The mechanisms of the electrons depletion \cite{Mamun1996,Dialynas2009,Mayout2012,Sahu2012,Ferdousi2015,Ferdousi2017},
in which maximum number of electrons are inserted onto the massive negative dust grains from the
background of DPM during the dust charging process, are ubiquitous
in space environments, viz., interstellar clouds \cite{Hossen2016b,Hossen2017}, F-ring of
Jupiter's magnetosphere \cite{Hossen2016a}, Earth polar mesosphere \cite{Hossen2016a},
Saturn \cite{Mayout2012,Sahu2012}, cometary tails \cite{Hossen2016a}, solar system \cite{Hossen2017}, and also in laboratory DPM.
The attachment of electrons with massive negatively charge dust grains is considered as electron depleted
DP (EDDP) medium (EDDPM). Shukla and Silin \cite{Shukla1992} examined DIAWs in an unmagnetized collisionless EDDPM.
Ferdousi \textit{et al.} \cite{Ferdousi2015} considered a two components plasma system containing inertialess positively charged ions and
inertial negatively charged dust grains to investigate DA-SHWs, and found that the model supports
both positive and negative electrostatic potentials. Mamun \textit{et al.} \cite{Mamun1996} considered
a two components EDDP model for investigation of the propagation of
nonlinear solitary pulses, and found that the height of the negative potential pulses
increases with the number density of ion and dust.
Sahu and Tribeche \cite{Sahu2012} studied electrostatic double-layers (DLs) in an unmagnetized EDDPM having inertial dust grains and inertiales ion,
and reported that both compressive and rarefactive DA-DLs are allowed by their plasma model.
Hossen \textit{et al.} \cite{Hossen2016b} examined DAWs in a multi-component EDDPM
having  inertialess non-thermal ions and inertial massive OP dust grains, and found that the
configuration of DA-DLs and DA-SWs is regourously modified by the existence of the positively charged dust grains.

The super-thermal or $\kappa$-distribution \cite{Vasyliunas1968,Uddin2015,Shahmansouri2012,Kourakis2011,Ferdousi2017,Sultana2011,Ahmed2018,Gill2010}
can describe the deviation, according to the values of the super-thermal parameter
$\kappa$ which manifests the presence of the external force fields or wave-particle
interactions, of plasma species from the thermal or Maxwellian distribution. The super-thermal or
$\kappa$-distribution exchanges with the Maxwellian distribution when $\kappa$ tends
to infinity, i.e., $\kappa\rightarrow\infty$, and $\kappa$-distribution
is normalizable for any kind of values of $\kappa$ rather than $\kappa>3/2$ \cite{Uddin2015,Shahmansouri2012,Kourakis2011,Sultana2011,Ahmed2018,Gill2010}.
Uddin \textit{et al.} \cite{Uddin2015} numerically analyzed the propagation of nonlinear electrostatic
positron-acoustic waves in a super-thermal plasma, and reported that the amplitude of the
electrostatic positive potential decreases with increasing value of $\kappa$.
Shahmansouri and Alinejad \cite{Shahmansouri2012} examined DA-SWs in a DPM having  super-thermal plasma species,
and found that the depth of the potential well decreases with the increase in the value of $\kappa$.
Kourakis and Sultana \cite{Kourakis2011} examined the speed of the DIA solitons in presence of the super-thermal
particles in a DPM.

The modulational instability (MI), energy localization, and energy redistribution of the carrier
waves are governed by the standard nonlinear Schr\"{o}dinger equation (NLSE) \cite{Sultana2011,Ahmed2018,Gill2010,Saini2018,C2,C2,C3,C4,C5}.
Sultana and Kourakis \cite{Sultana2011} examined the electron-acoustic (EA) envelope solitons in a plasma medium having
super-thermal electrons, and found that the stable domain of EA waves decreases with increasing $\kappa$.
Ahmed \textit{et al.} \cite{Ahmed2018} reported IAWs in a four-components plasma medium, and
highlighted that the critical wave number ($k_c$) increases with a decrease in the value of $\kappa$.
Saini and Kourakis \cite{Saini2018} demonstrated the MI of the DAWs a DPM having super-thermal ions,
and obtained that excess super-thermality of the ions leads a narrower envelope solitons.

Recently, Hossen \textit{et al.} \cite{Hossen2016a} considered three components plasma model having
inertial OP dust grains and non-extensive electrons to investigate the propagation of the DA-SWs.
In this paper, we want to develop sufficient extension of previous published work \cite{Hossen2016a}
by considering a real and novel four components EDDP model having inertial OP dust grains and inertialess
iso-thermal negatives ions and super-thermal positive ions to examine the MI of DAWs and formation of DA rogue
waves (DARWs).

The layout of the paper is as follows: The governing equations are given in Section \ref{1sec:Basic Equations}.
The derivation of the standard NLSE is devoted in Section \ref{1sec:Derivation of the NLSE}.
MI and rogue waves are presented in Section \ref{1sec:Modulational instability and rogue waves}.
Results and discussion are provided in Section \ref{1sec:Results and discussion}.
A brief conclusion is demonstrated in Section \ref{1sec:Conclusion}.
\section{Basic Equations}
\label{1sec:Basic Equations}
We consider a four components plasma medium consisting of inertial positively and negatively
charged massive dust grains, and inertialess positive and  negative ions. At equilibrium, the
quasi-neutrality condition can be expressed as $Z_+n_{+0}+n_{20}=Z_-n_{-0}+n_{10}$; where
$n_{+0}$, $n_{10}$, $n_{-0}$, and $n_{20}$ are, respectively, the equilibrium number densities of positive dust grains, positive ions,
negative dust grains, and negative ions. Now, after normalization the set of the equations can be written as
\begin{eqnarray}
&&\hspace*{-1.3cm}\frac{\partial n_{-}}{\partial t}+\frac{\partial}{\partial x}(n_{-}u_{-})=0,
\label{1eq:1}\\
&&\hspace*{-1.3cm}\frac{\partial u_-}{\partial t}+u_-\frac{\partial u_-}{\partial x}=\lambda_1\frac{\partial\phi}{\partial x},
\label{1eq:2}\\
&&\hspace*{-1.3cm}\frac{\partial n_+}{\partial t}+\frac{\partial}{\partial x}(n_+u_+)=0,
\label{1eq:3}\\
&&\hspace*{-1.3cm}\frac{\partial u_+}{\partial t}+u_+\frac{\partial u_+}{\partial x}=-\frac{\partial\phi}{\partial x},
\label{1eq:4}\\
&&\hspace*{-1.3cm}\frac{\partial^2\phi}{\partial x^2}=\lambda_2n_1-\lambda_3n_2+(1-\lambda_2+\lambda_3)n_--n_+,
\label{1eq:5}
\end{eqnarray}
where $n_{-}$ and $n_{+}$  are the negative and positive dust grains number density normalized
by their equilibrium value $n_{-0}$ and $n_{+0}$, respectively; $u_{-}$ and $u_{+}$ are
the negative and positive dust fluid speed normalized by wave speed $C_{+}=(Z_+K_BT_-/m_+)^{1/2}$
(with $T_-$ being the temperature negative ion, $m_+$ being the positive dust mass,
and $K_B$ being the Boltzmann constant); $\phi$ is the electrostatic wave potential
normalized by $K_BT_-/e$ (with $e$ being the magnitude of single electron charge); the
time and space variables are normalized by $\omega_{p+}^{-1}=(m_+/4\pi e^2 Z_+^2n_{+0})^{1/2}$
and $\lambda_{D+}=(k_BT_-/4\pi e^2 Z_+n_{+0})^{1/2}$, respectively; and $\lambda_1=Z_-m_+/Z_+m_-$,
$\lambda_2=Z_1n_{10}/Z_+n_{+0}$, and $\lambda_3=Z_2n_{20}/Z_+n_{+0}$. We have considered $m_-> m_+$, $Z_->Z_+$, and $n_{-0}>n_{+0}$ for our numerical analysis.
Now, the expression for  positive ion number density obeying $\kappa$-distribution is given by \cite{Ahmed2018}
\begin{eqnarray}
&&\hspace*{-1.3cm}n_1=\left[1 +\frac{\lambda_4\phi}{\kappa-3/2}\right]^{-\kappa+1/2},
\label{1eq:6}
\end{eqnarray}
where $\lambda_4 =T_-/T_+$ (with $T_+$ being the temperature positive ion) and $T_+>T_-$.
The super-thermality of the light positive ion can be represented by the  parameter $\kappa$.
The expression for negative ion number density obeying iso-thermal Maxwelliann distribution is given by
\begin{eqnarray}
&&\hspace*{-1.3cm}n_2=\mbox{exp}(\phi).
\label{1eq:7}
\end{eqnarray}
Now, by substituting Eqs. \eqref{1eq:6} and \eqref{1eq:7} into Eq. \eqref{1eq:5}, and expanding up to third order of $\phi$, we get
\begin{eqnarray}
&&\hspace*{-1.3cm}\frac{\partial^2\phi}{\partial x^2} + n_{+}+\lambda_3= \lambda_2 +(1-\lambda_2+\lambda_3)n_{-}
\nonumber\\
&&\hspace*{1.0cm}+ T_1\phi + T_2\phi^2 + T_3\phi^3 +\cdot\cdot\cdot,
\label{1eq:8}
\end{eqnarray}
where
\begin{eqnarray}
&&\hspace*{-1.3cm}T_1=\frac{\lambda_2(2\kappa-3)+\lambda_3\lambda_4(2\kappa-1)}{(2\kappa-3)},
\nonumber\\
&&\hspace*{-1.3cm}T_2=\frac{\lambda_2(2\kappa-3)^2-\lambda_3\lambda_4^2(2\kappa-1)(2\kappa+1)}{2(2\kappa-3)^2},
\nonumber\\
&&\hspace*{-1.3cm}T_3=\frac{\lambda_2(2\kappa-3)^3+\lambda_3\lambda_4^3(2\kappa-1)(2\kappa+1)(2\kappa+3)}{6(2\kappa-3)^3}.
\nonumber\
\end{eqnarray}
\section{Derivation of the NLSE}
\label{1sec:Derivation of the NLSE}
The reductive perturbation method is applicable to derive the standard NLSE as well as to study the MI of the DAWs in a
four components electrons depleted dusty plasma. The stretched co-ordinates, to develop a standard NLSE, can be written as
\begin{eqnarray}
&&\hspace*{-1.3cm}\xi={\epsilon}(x-v_g t),
\label{1eq:9}\\
&&\hspace*{-1.3cm}\tau={\epsilon}^2 t,
\label{1eq:10}
\end{eqnarray}
where $v_g$ is the group velocity and $\epsilon$ ($\epsilon\ll 1$) is a small parameter which measures
the nonlinearity of the plasma medium. Then, the dependent variables can be written as
\begin{eqnarray}
&&\hspace*{-1.3cm}n_-=1+\sum_{m=1}^{\infty}\epsilon^{m}\sum_{l=-\infty}^{\infty} n_{-l}^{(m)}(\xi,\tau)~\mbox{exp}[i l(kx-\omega t)],
\label{1eq:11}\\
&&\hspace*{-1.3cm}u_-=\sum_{m=1}^{\infty}\epsilon^{m}\sum_{l=-\infty}^{\infty} u_{-l}^{(m)}(\xi,\tau)~\mbox{exp}[i l(kx-\omega t)],
\label{1eq:12}\\
&&\hspace*{-1.3cm}n_+=1+\sum_{m=1}^{\infty}\epsilon^{m}\sum_{l=-\infty}^{\infty} n_{+l}^{(m)}(\xi,\tau)~\mbox{exp}[i l(kx-\omega t)],
\label{1eq:13}\\
&&\hspace*{-1.3cm}u_+=\sum_{m=1}^{\infty}\epsilon^{m}\sum_{l=-\infty}^{\infty} u_{+l}^{(m)}(\xi,\tau)~\mbox{exp}[i l(kx-\omega t)],
\label{1eq:14}\\
&&\hspace*{-1.3cm}\phi=\sum_{m=1}^{\infty}\epsilon^{m}\sum_{l=-\infty}^{\infty} \phi_{l}^{(m)}(\xi,\tau)~\mbox{exp}[i l(kx-\omega t)].
\label{1eq:15}\
\end{eqnarray}
For the above consideration, the derivative operators can be recognized as
\begin{eqnarray}
&&\hspace*{-1.3cm}\frac{\partial}{\partial t}\rightarrow\frac{\partial}{\partial t}-\epsilon v_g \frac{\partial}{\partial\xi}
+\epsilon^2\frac{\partial}{\partial\tau},
\label{1eq:16}\\
&&\hspace*{-1.3cm}\frac{\partial}{\partial x}\rightarrow\frac{\partial}{\partial x}+\epsilon\frac{\partial}{\partial\xi}.
\label{1eq:17}
\end{eqnarray}
Now, by substituting Eqs. \eqref{1eq:9}-\eqref{1eq:17}  into Eqs. \eqref{1eq:1}-\eqref{1eq:4}, and \eqref{1eq:8}, and
collecting the terms containing $\epsilon$, the first order ($m=1$ with $l=1$)  reduced equations can be written as
\begin{eqnarray}
&&\hspace*{-1.3cm}n_{-1}^{(1)}=-\frac{\lambda_1 k^2}{\omega^2}\phi_1^{(1)},
\label{1eq:18}\\
&&\hspace*{-1.3cm}u_{-1}^{(1)}=-\frac{\lambda_1 k}{\omega}\phi_1^{(1)},
\label{1eq:19}\\
&&\hspace*{-1.3cm}n_{+1}^{(1)}=\frac{k^2}{\omega^2}\phi_1^{(1)},
\label{1eq:20}\\
&&\hspace*{-1.3cm}u_{+1}^{(1)}=\frac{k }{\omega}\phi_1^{(1)},
\label{1eq:21}\
\end{eqnarray}
these relation provides the dispersion relation for DAWs
\begin{eqnarray}
&&\hspace*{-1.3cm}\omega^2=\frac{k^2+\lambda_1 k^2-\lambda_1\lambda_2 k^2+\lambda_1\lambda_3 k^2}{k^2+T_1}.
\label{1eq:22}
\end{eqnarray}
The second-order ($m=2$ with $l=1$) equations are given by
\begin{eqnarray}
&&\hspace*{-1.3cm}n_{-1}^{(2)}=-\frac{\lambda_1 k^2}{\omega^2}\phi_1^{(2)}-\frac{2i\lambda_1k(v_gk-\omega)}{\omega^3}\frac{\partial \phi_1^{(1)}}{\partial\xi},
\label{1eq:23}\\
&&\hspace*{-1.3cm}u_{-1}^{(2)}=-\frac{\lambda_1 k}{\omega}\phi_1^{(2)}-\frac{i\lambda_1(v_gk-\omega)}{\omega^2}\frac{\partial \phi_1^{(1)}}{\partial\xi},
\label{1eq:24}\\
&&\hspace*{-1.3cm}n_{+1}^{(2)}=\frac{k^2}{\omega^2}\phi_1^{(2)}+\frac{2ik(v_gk-\omega)}{\omega^3}\frac{\partial \phi_1^{(1)}}{\partial\xi},
\label{1eq:25}\\
&&\hspace*{-1.3cm}u_{+1}^{(2)}=\frac{k}{\omega}\phi_1^{(2)}+\frac{i(v_gk-\omega)}{\omega^2}\frac{\partial \phi_1^{(1)}}{\partial\xi},
\label{1eq:26}\
\end{eqnarray}
with the compatibility condition
\begin{eqnarray}
&&\hspace*{-1.3cm}v_g=\frac{\omega+\omega\lambda_1-\omega\lambda_1\lambda_2+\omega\lambda_1\lambda_3-\omega^3}{k+k\lambda_1-k\lambda_1\lambda_2+k\lambda_1\lambda_3                            }.
\label{1eq:27}\
\end{eqnarray}
The coefficients of $\epsilon$ for $m=2$ with $l=2$ provide the second
order harmonic amplitudes which are found to be proportional to $|\phi_1^{(1)}|^2$
\begin{eqnarray}
&&\hspace*{-1.3cm}n_{_2}^{(2)}=T_4|\phi_1^{(1)}|^2,
\label{1eq:28}\\
&&\hspace*{-1.3cm}u_{-2}^{(2)}=T_5 |\phi_1^{(1)}|^2,
\label{1eq:29}\\
&&\hspace*{-1.3cm}n_{+2}^{(2)}=T_6|\phi_1^{(1)}|^2,
\label{1eq:30}\\
&&\hspace*{-1.3cm}u_{+2}^{(2)}=T_7 |\phi_1^{(1)}|^2,
\label{1eq:31}\\
&&\hspace*{-1.3cm}\phi_{2}^{(2)}=T_8 |\phi_1^{(1)}|^2,
\label{1eq:32}\
\end{eqnarray}
where
\begin{eqnarray}
&&\hspace*{-1.3cm}T_4=\frac{\lambda_1k^2(3\lambda_1k^2-2\omega^2T_8)}{2\omega^4},
\nonumber\\
&&\hspace*{-1.3cm}T_5=\frac{\lambda_1k(\lambda_1k^2-2\omega^2T_8)}{2\omega^3},
\nonumber\\
&&\hspace*{-1.3cm}T_6=\frac{k^2(2\omega^2T_8+3k^2)}{2\omega^4},
\nonumber\\
&&\hspace*{-1.3cm}T_7=\frac{k(2\omega^2T_8+k^2)}{2\omega^3},
\nonumber\\
&&\hspace*{-1.3cm}T_8=\frac{2T_2\omega^4-3k^4(1+\lambda_2\lambda_1^2-\lambda_1^2-\lambda_3\lambda_1^2)}{F1-2\omega^4(4k^2+T_1)},
\nonumber\
\end{eqnarray}
where $F1=2\omega^2k^2(1+\lambda_1-\lambda_1\lambda_2+\lambda_1\lambda_3)$. Now, we consider the expression for ($m=3$ with $l=0$) and ($m=2$ with $l=0$),
which leads the zeroth harmonic modes. Thus, we obtain
\begin{eqnarray}
&&\hspace*{-1.3cm}n_{-0}^{(2)}=T_{9}|\phi_1^{(1)}|^2,
\label{1eq:33}\\
&&\hspace*{-1.3cm}u_{-0}^{(2)}=T_{10}|\phi_1^{(1)}|^2,
\label{1eq:34}\\
&&\hspace*{-1.3cm}n_{+0}^{(2)}=T_{11}\phi_1^{(1)}|^2,
\label{1eq:35}\\
&&\hspace*{-1.3cm}u_{+0}^{(2)}=T_{12}|\phi_1^{(1)}|^2,
\label{1eq:36}\\
&&\hspace*{-1.3cm}\phi_0^{(2)}=T_{13}|\phi_1^{(1)}|^2,
\label{1eq:37}\
\end{eqnarray}
where
\begin{eqnarray}
&&\hspace*{-1.3cm}T_{9}=\frac{\lambda_1^2k^2(\omega+2kv_g)-T_{13}\lambda_1\omega^2}{v_g^2\omega^3},
\nonumber\\
&&\hspace*{-1.3cm}T_{10}=\frac{\lambda_1(\lambda_1k^2- \omega^2T_{13})}{v_g\omega^2},
\nonumber\\
&&\hspace*{-1.3cm}T_{11}=\frac{\omega(k^2+\omega^2 T_{13})+2v_gk^3}{v_g^2\omega^3},
\nonumber\\
&&\hspace*{-1.3cm}T_{12}=\frac{k^2+\omega^2 T_{13}}{v_g\omega^2},
\nonumber\\
&&\hspace*{-1.3cm}T_{13}=\frac{2v_g(v_g\omega^3T_2-k^3)-\omega k^2+(1-\lambda_2+\lambda_3)\times F2}{\omega^3+\lambda_1\omega^3-\lambda_1\lambda_2\omega^3+\lambda_1\lambda_3\omega^3-T_1\omega^3v_g^2},
\nonumber\
\end{eqnarray}
where $F2=\omega k^2\lambda_1^2+2v_g k^3 \lambda_1^2$. Finally, the third harmonic modes ($m=3$) and ($l=1$) with the help of Eqs. \eqref{1eq:9}-\eqref{1eq:37}, give a set of equations, which can be reduced to the following NLSE:
\begin{eqnarray}
&&\hspace*{-1.3cm}i\frac{\partial \Phi}{\partial\tau}+P\frac{\partial^2\Phi}{\partial\xi^2}+Q\mid\Phi\mid^2\Phi=0,
\label{1eq:38}
\end{eqnarray}
where $\Phi=\phi_1^{(1)}$ for simplicity. In Eq. \eqref{1eq:38}, $P$ is the dispersion coefficient which can be written as
\begin{eqnarray}
&&\hspace*{-1.3cm}P=\frac{3v_g(v_g k-\omega)}{2\omega k},
\nonumber\
\end{eqnarray}
and also $Q$ is the nonlinear coefficient which can be written as
\begin{eqnarray}
&&\hspace*{-1.3cm}Q=\frac{3\omega^3T_3+2\omega^3T_2(T_8+T_{13})-F3}{2k^2+2\lambda_1 k^2-2\lambda_1\lambda_2 k^2+2\lambda_1\lambda_3 k^2}
\nonumber\
\end{eqnarray}
where
\begin{eqnarray}
&&\hspace*{-1.3cm}F3=\omega k^2(T_6+T_{11})+2k^3(T_7+T_{12})
\nonumber\\
&&\hspace*{-0.5cm}+(\omega\lambda_1 k^2-\omega\lambda_1\lambda_2k^2+\omega\lambda_1\lambda_3 k^2)(T_4+T_9)
\nonumber\\
&&\hspace*{-0.5cm}+(2\lambda_1 k^3-2\lambda_1\lambda_2k^3+2\lambda_1\lambda_3 k^3)(T_5+T_{10}).
\nonumber\
\end{eqnarray}
It may be noted here that both $P$ and $Q$ are function of various
plasma parameters such as $k$, $\lambda_1$, $\lambda_2$, $\lambda_3$, $\lambda_4$, and $\kappa$.
So, all the plasma parameters are used to maintain
the nonlinearity and the dispersion properties of the DPM.
\begin{figure}[t!]
\centering
\includegraphics[width=80mm]{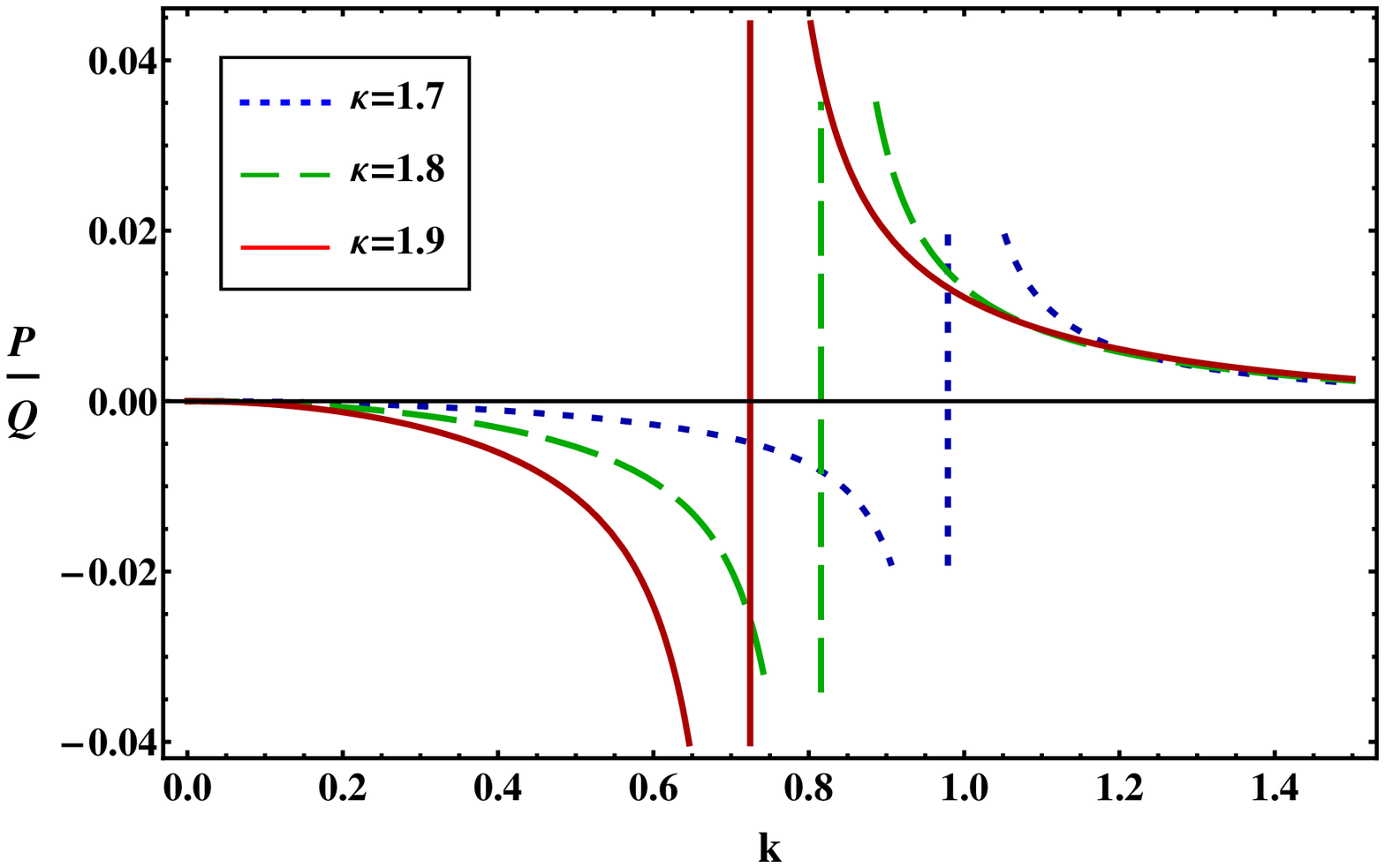}
\caption{Plot of $P$ vs $k$ for various values of $\kappa$ when $\lambda_1=0.7$, $\lambda_2=2.0$, $\lambda_3=1.5$, and $\lambda_4=0.4$.}
\label{1Fig:F1}
\vspace{0.8cm}
\includegraphics[width=80mm]{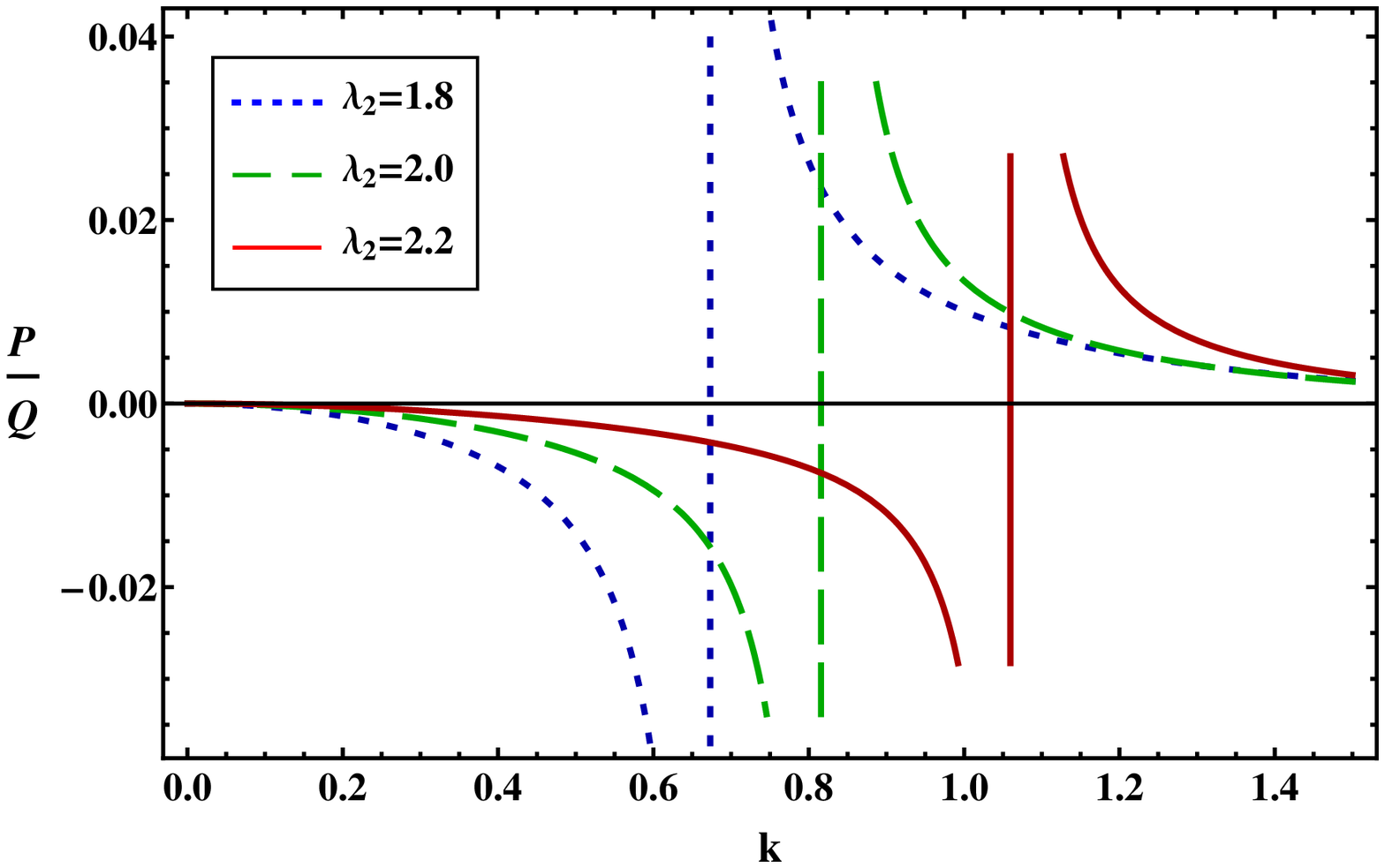}
\caption{Plot of $Q$ vs $k$ for various values of $\lambda_2$ when $\lambda_1=0.7$, $\lambda_3=1.5$, $\lambda_4=0.4$, and $\kappa=1.8$.}
 \label{1Fig:F2}
\end{figure}
\begin{figure}[t!]
\centering
\includegraphics[width=80mm]{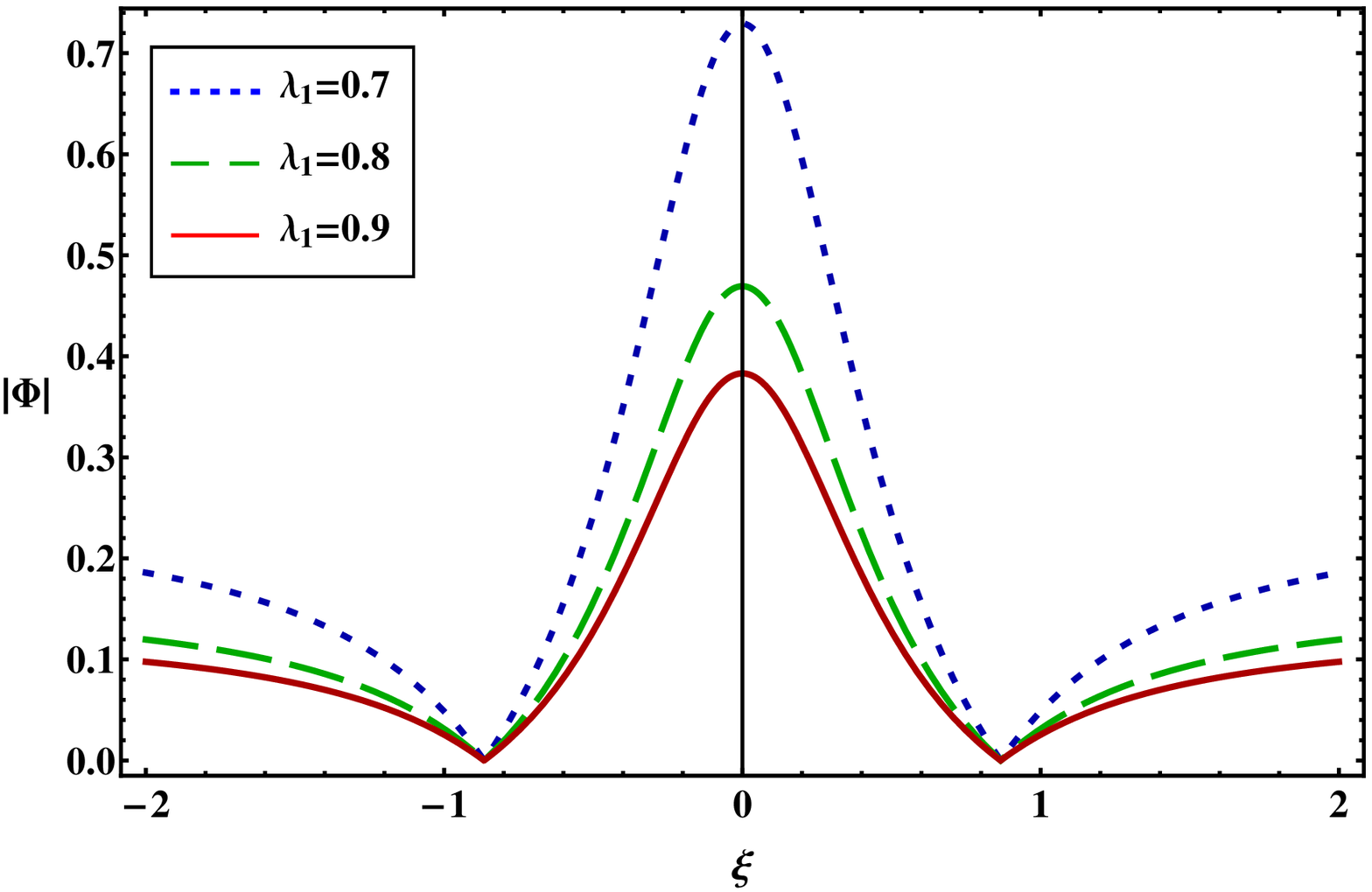}
\caption{Plot of $|\Phi|$ vs $\xi$ for various values of  $\lambda_1$, when $\lambda_2=2.0$, $\lambda_3=1.5$, $\lambda_4=0.4$, and $\kappa=1.8$.}
\label{1Fig:F3}
\vspace{0.8cm}
\includegraphics[width=80mm]{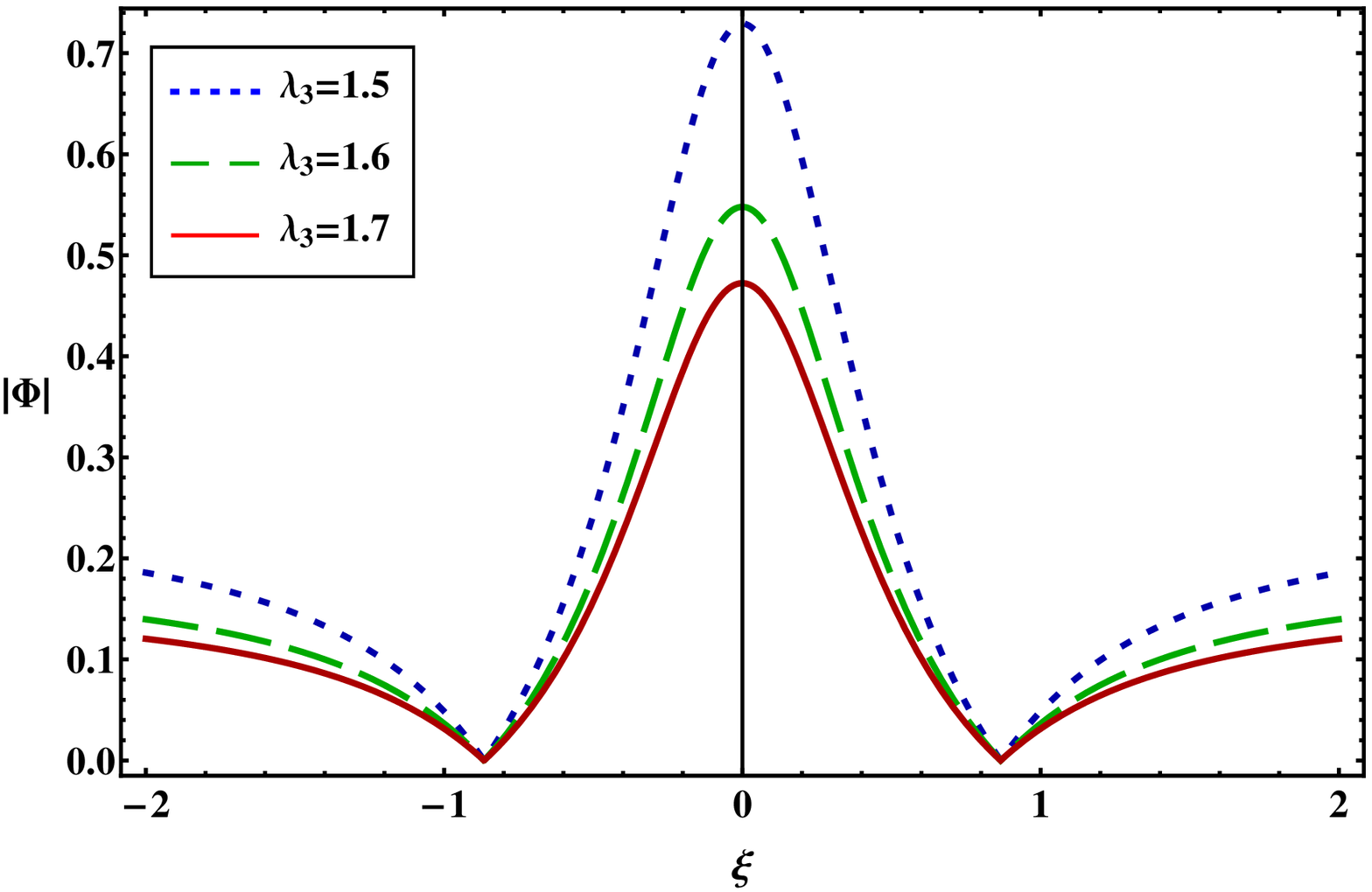}
\caption{Plot of $|\Phi|$ vs $\xi$ for various values of $\lambda_3$, when $\lambda_1=0.7$, $\lambda_2=2.0$, $\lambda_4=0.4$, and $\kappa=1.8$.}
 \label{1Fig:F4}
\end{figure}
\section{Modulational instability and rogue waves}
\label{1sec:Modulational instability and rogue waves}
The nonlinear property of the plasma medium as well as the formation and stability conditions of the DAWs
in an EDDPM can be organized according to the sign of dispersive ($P$) and nonlinear
($Q$) coefficients of the standard NLSE \eqref{1eq:38}. The sign of the dispersive coefficient is always negative for
any kinds of wave number $k$ in $P$ vs $k$ curve while the sign of the nonlinear coefficient is positive for small
values of $k$ and negative for large values of $k$ (figure is not included). Both $P$ and $Q$ have same
sign (i.e., positive or negative) then they reflect a modulationally unstable parametric regime (i.e., $P/Q>0$) whereas both $P$ and $Q$ have
opposite sign (i.e., positive and negative) then they reflect a modulationally stable parametric regime (i.e., $P/Q<0$) for the DAWs in presence
of the external perturbation. The stable and unstable parametric regimes are differentiated by a vertical line at
which $P/Q\rightarrow\infty$ (i.e., $Q=0$ and because $P$ is always negative), and the wave number for which $P/Q\rightarrow\infty$
is known as critical waves number ($k_c$) \cite{C6,C7,C8,C9,C10}.

The modulationally unstable parametric regime of the DAWs allows to generate highly energetic and mysterious DARWs associated with DAWs,
and the governing equation of the puzzling DARWs can be written as \cite{Akhmediev2009a,Ankiewiez2009b}
\begin{eqnarray}
&&\hspace*{-1.3cm}\Phi(\xi,\tau)= \sqrt{\frac{2P}{Q}} \left[\frac{4(1+4iP\tau)}{1+16P^2\tau^2+4\xi^2}-1 \right]\mbox{exp}(2iP\tau).
\label{1:eq39}
\end{eqnarray}
We have numerically analyzed Eq. \eqref{1:eq39} in Figs. \ref{1Fig:F3} and \ref{1Fig:F4} to understand
how the nonlinear properties of a four components EDDPM as well as the configuration of the
DARWs associated DAWs have been changed by the various plasma parameters.
\section{Results and discussion}
\label{1sec:Results and discussion}
The variation of $P/Q$ with $k$ for different values of $\kappa$ as well as the stable and unstable
parametric regimes of the DAWs can be observed in Fig. \ref{1Fig:F1}, and it is obvious from this
figure that (a) DAWs become modulationally stable for small values of $k$ while unstable for large
values of $k$; (b) the $k_c$ decreases with the increase in the value of the $\kappa$. Figure \ref{1Fig:F2}
indicates the effects of the number density as well as the charge state of the positive dust grains and
negative ions (via $\lambda_2=Z_1n_{10}/Z_+n_{+0}$) to recognize the stable and unstable parametric regimes
of the DAWs, and it can be seen from this figure that (a) the modulationally stable parametric regime increases with $\lambda_2$;
(b) the modulationally stable parametric regime increases with an increase in the value of negative ion number density ($n_{10}$)
while decreases with an increase in the value of the positive dust grains number density ($n_{+0}$) for a constant value of $Z_1$ and $Z_+$;
(c) the modulationally unstable (stable) parametric regime of the DAWs increases with $Z_+$ ($Z_1$) for a fixed negative ion and positive dust
number density.

We have numerically analyzed first-order rogue waves [by using Eq. \eqref{1:eq39}] in Figs. \ref{1Fig:F3}
and \ref{1Fig:F4}. Figure \ref{1Fig:F3} highlights the effects of mass of the positive and negative ions
as well as their charge states (via $\lambda_2=Z_-m_+/Z_+m_-$) in recognizing the shape of the DARWs in
an electron depleted four components EDDPM, and it can be manifested from this figure that (a) the
increase in the value of $\lambda$ does not only cause to change the amplitude of the DARWs but also causes
to change the width of the DARWs; (b) the amplitude and width of the DARWs decrease with the increase in the
value of $\lambda_2$; (c) actually, the nonlinearity of the plasma medium as well as the amplitude of the DARWs increases with
increasing $m_-$ while the nonlinearity as well as the amplitude of the DARWs decreases with increasing $m_+$
for a fixed value of $Z_+$ and $Z_-$. Figure \ref{1Fig:F4} reflects how the number density and charge state of the
light positive ion and heavy negative dust grains (via $\lambda_3=Z_2n_{20}/Z_+n_{+0}$) contribute to generate
highly energetic rogue waves in a four components EDDPM. The amplitude and width of the
electrostatic DARWs associated with DAWs decreases with an increase in the value of positive ion charge state and number density
while increases with an increase in the value of positive dust charge state and number density. The physics of this result is that
the nonlinearity of a four components electron depleted plasma medium increases (decreases) with positive dust grain (positive ion) number density
as well as with positive dust grain (positive ion) charge state.
\section{Conclusion}
\label{1sec:Conclusion}
In this present article, we have examined the modulationally stable and unstable parametric regimes of DAWs, and the DARWs in the unstable parametric regime
of DAWs by employing standard NLSE in a four components EDDPM. The numerical analysis can explain the
dependency of the stability conditions as well as the configuration of DARWs associated with DAWs in the modulationally unstable parametric regime.
To conclude, the results obtained from this investigation should be useful in understanding the basic properties of these
rouge waves predicted to be formed in electron depleted unmagnetized opposite polarity dusty plasma systems like mesospherere, F-rings of saturn, and
cometary atmosphere, etc.
\section*{Acknowledgements}
A. Mannan and S. Sultana gratefully acknowledge the support from the Alexander von Humboldt Foundation for via their Postdoctoral Fellowship.

\end{document}